\documentclass[
    aps,
    prb,
    reprint,
    superscriptaddress,
    amsmath,
    amssymb,
    floatfix,
    showpacs
]{revtex4-1}

\usepackage[T1]{fontenc}
\usepackage[utf8]{inputenc}
\usepackage{lmodern}
\usepackage{microtype}

\usepackage{amsfonts,amsthm,mathtools}
\usepackage{bm}
\usepackage{graphicx}
\usepackage{xcolor}
\usepackage{booktabs}
\usepackage[colorlinks=true,linkcolor=blue,citecolor=blue,urlcolor=blue]{hyperref}

\numberwithin{equation}{section}

\theoremstyle{plain}

\theoremstyle{definition}

\newcommand{\diag}{\operatorname{diag}}
\newcommand{\ii}{\mathrm{i}}

\newcommand{\dd}{\mathrm{d}}

\newcommand{\one}{\mathbf{1}}
\newcommand{\zero}{\mathbf{0}}

\newcommand{\T}{\mathbf{T}}
\newcommand{\tmat}{\mathbf{t}}
\newcommand{\rmat}{\mathbf{r}}

\newcommand{\smat}{\mathbf{s}}
\newcommand{\kmat}{\mathbf{k}}
\newcommand{\zmat}{\mathbf{z}}
\newcommand{\Jmat}{\mathbf{J}}
\newcommand{\Kmat}{\mathbf{K}}
\newcommand{\Amat}{\mathbf{A}}
\newcommand{\Nmat}{\mathbf{N}}

\newcommand{\Lmat}{\mathbf{L}}
\newcommand{\Mmat}{\mathbf{M}}
\newcommand{\Xmat}{\mathbf{X}}
\newcommand{\Smat}{\mathbf{S}}
\newcommand{\Rmat}{\mathbf{R}}

\newcommand{\bsig}{\boldsymbol{\sigma}}
\newcommand{\balp}{\boldsymbol{\alpha}}
\newcommand{\bbet}{\boldsymbol{\beta}}

\newcommand{\bPsi}{\boldsymbol{\Psi}}
\newcommand{\bPhi}{\boldsymbol{\Phi}}

\newcommand{\ba}{\mathbf{a}}
\newcommand{\bb}{\mathbf{b}}
\newcommand{\bc}{\mathbf{c}}
\newcommand{\bd}{\mathbf{d}}

\newcommand{\ham}{\mathcal{H}}

\newcommand{\vct}[1]{\boldsymbol{#1}}

\begin{document}

\title{Geometric aspects of spin transport in magnetic multilayers}
\author{Valentin Fadeev}
\date{7 Jun 2026}
	
	\begin{abstract}
		We discuss spin-dependent transfer-matrix formalism applied to magnetic multilayers in geometric terms. Starting from the stationary Schr\"odinger equation rewritten as a first-order spatial evolution problem, 
		we interpret the transfer matrix as a path-ordered exponential and relate its matching-matrix construction to a noncompact group constraint. We then connect the induced M\"obius action on reflection matrices to an 
		Iwasawa decomposition, identify Weyl-chamber variables as the minimal noncompact transport invariants, and show how torque-related spin structures arise from compact factor and commutator contributions. A sequence of multilayer examples illustrates the transition from pure spin filtering to controlled spin--orbit mixing and the resulting deformation of Weyl-chamber trajectories. We finally comment on the extension to higher-dimensional internal spaces relevant to orbital transport and realistic calculations.
	\end{abstract}
	
	
	\maketitle
	
	\section{Introduction}
	The study of magnetic multilayers both heralded and was spurred by the advent of spintronics in the recent decades. From magnetic tunnel junctions \cite{Tsymbal2026MagneticTunnelJunctions} to synthetic antiferromagnets \cite{Duine2018SyntheticAntiferromagneticSpintronics} and beyond, stacks of thin films of materials are the building blocks and archetypal models when it comes to investigating and harnessing spin-resolved electron transport phenomena. Engineering applications include large areas of interest, such as non-volatile memories (MRAM) \cite{Yang_2025} and terahertz radiation emitters \cite{Kampfrath2013,Jungfleisch2018}. Furthermore, recent progress in nanofabrication and material characterisation makes fully phase-coherent treatment of the transport quantitatively relevant, rather than being an illustrative simplification. In this article we turn our attention to phase-coherent spin-resolved transport in magnetic multilayers and attempt to establish a language for characterising their action on the input current. In particular, spin injection and spin filtering efficiency are important parameters of spintronic devices. We will see how those are related to noncompact factors of the transformations realised by transfer matrices.
	
	The transfer matrix formalism is an established and well-understood tool in the study of layered media. Although, as a sole numerical method, it may be limited in its utility for realistic calculations by well-known stability issues, it remains extremely valuable for exposing the internal structure of transport problems in terms of a reduced set of independent variables. In the spinless setting, the geometric viewpoint on transfer matrices has been extensively developed and forms a common language across one-dimensional quantum scattering and linear optics \cite{SANCHEZSOTO2012191,Yonte:02,Monzon:01}. By contrast, applications to spin-resolved transport appear to be much less systematically organised. 
	
	The purpose of this work is not to introduce a new numerical method for
	spin-dependent transport, nor to replace standard scattering or Green's function
	formulations. Rather, we reorganise the coherent transfer matrix description of
	magnetic multilayers in geometric terms. This reveals a natural projective
	Möbius action on reflection matrices, an Iwasawa interpretation in terms of
	rotations, scaling, and shear, a concise Weyl-chamber description of radial
	spin filtering, and a real observable-space Spin--Mueller--Jones lift acting on charge and spin current components.
	The use of Cartan-type radial variables for transfer matrices is well known in mesoscopic quantum transport and random-matrix theory \cite{RevModPhys.69.731,CASELLE200441}. The present work applies this geometric viewpoint to the spin-resolved case and uses the singular values of the  reflection matrix as Weyl-chamber coordinates for spin filtering.
	Similarly, the Mueller--Jones construction is standard in polarisation optics \cite{Jones:41,Schmieder:69,aiello2006linearalgebramuellercalculus}. Here we use its algebraic form as an observable-space lift of the spinful transfer matrix action, producing a real tensor acting on charge and spin current components. We are not aware of a previous application of this specific direction--spin Mueller--Jones tensor to coherent spin transport in magnetic multilayers.
    To motivate the use of this mathematical framework, we apply it to a sequence of model structures relevant for applications in spintronics. These examples provide a simple characterisation of the action of magnetic multilayers, show how torque emerges within the formalism, and clarify the role of spin-orbit coupling in a unified language. A central outcome is that an apparently high-dimensional problem, not obviously amenable to direct visualisation, reduces to an interpretable picture in terms of trajectories in Weyl-chamber variables. We conclude by outlining a possible extension of the formalism to higher-dimensional internal spaces relevant to orbital transport and realistic numerical calculations.
	\section{Transfer matrix as a spatially ordered exponential}	
	In this section we introduce the transfer matrix as a path-ordered exponential. This representation serves to illustrate how the concept naturally follows from casting the stationary Schr\"odinger equation to the form suitable for studying scattering in layered systems.
	
	\subsection{Stationary transport as spatial evolution}
	\label{subsec:spatial-evolution}
	Consider the stationary Schr\"odinger equation
	\begin{equation}
		\ham\psi(x) = E\psi(x). \label{eq:stationary_schroedinger}
	\end{equation}
	Define
	\begin{equation*}
		\bPhi(x) = \begin{pmatrix}
			\psi(x) \\
			\psi'(x)
		\end{pmatrix}.
	\end{equation*}
	Now we can rewrite \eqref{eq:stationary_schroedinger} as a first-order system in space
	\begin{equation}
		\partial_x \bPhi(x) = \mathcal{K}(x)\bPhi(x) \label{eq:first_order}
	\end{equation}
	The transfer matrix is then given by the path-ordered exponential \cite{BoonsermVisser2010,TorresDelCastillo2006}
	\begin{equation}
		\T(x_2,x_1) = \mathcal{P}\exp\!\left(\int_{x_1}^{x_2}\dd x\,\mathcal{K}(x)\right). \label{eq:t_path_ordered_exponential}
	\end{equation}
	The expression \eqref{eq:t_path_ordered_exponential} is reminiscent of the exponential representation of the scattering matrix
	\begin{equation}
		\Smat = \mathcal{T}\exp\!\left(-i\int\dd t\,\mathcal{H_I}\right),
	\end{equation}
	where $\mathcal{T}$ is the time-ordering operator. Thus while $\Smat$ encodes evolution in time, $\T$ encodes ordered propagation in space.
	
	We now illustrate the general spatial-evolution viewpoint introduced above, as applied to the case of a simple spin-resolved effective Hamiltonian relevant to magnetic multilayers. Consider
	\begin{equation*}
		\ham = -\frac{\hbar^2}{2m}\partial^2_x + V(x) + \Delta(x)\vct m(x)\cdot\bsig+\ham_{\text{SOC}},
	\end{equation*}
	where $V$ is the scalar potential, $\Delta$ is exchange splitting, $\vct m$ is the direction of the magnetisation. $\ham_{\text{SOC}}$ is a SOC term, which in a one-dimensional model can be parametrised as follows $\ham_{\text{SOC}}=-\ii\,\vct A\cdot\bsig\,\partial_x .$, where $\vct A$ is an effective SOC vector, for example arising from interfacial Rashba-type coupling. The stationary equation can then be rearranged as follows
	\begin{equation*}
		\psi'' = -\frac{2m}{\hbar^2}\left[-i\left(\vct A\cdot\bsig\right)\psi' + \left(V-E+\Delta\vct m\cdot\bsig\right)\psi\right],
	\end{equation*}
	which can be immediately cast into the form of \eqref{eq:first_order} with
	\begin{equation*}
		\mathcal{K}(x) = 
		\begin{pmatrix}
			\zero & \one \\
			-\frac{2m}{\hbar^2}\left(V-E+\Delta\vct m\cdot\bsig\right) & \frac{2m}{\hbar^2}\ii\,\vct A\cdot\bsig
		\end{pmatrix}
	\end{equation*}
	Now defining
	\begin{equation}
\mathbf{\Omega}_{\Phi} =
\begin{pmatrix}
	\zero & \one\\
	-\one & \zero
\end{pmatrix}, \label{eq:wronskian}
\end{equation}
	with $\Phi$ in the subscript to serve as a reminder that we are working in the derivative basis. In the absence of the first-derivative SOC term, the generator preserves \eqref{eq:wronskian} in the sense that
	\[
	\mathcal K^\dagger \mathbf{\Omega}_{\Phi} + \mathbf{\Omega}_{\Phi} \mathcal K = \zero.
	\]
	With SOC terms linear in momentum, the conserved current form is modified by the corresponding velocity operator. In either case, the first-order formulation preserves a bilinear form associated with current conservation. In a flux-normalised right/left-moving channel basis this conservation law becomes the pseudo-unitary constraint on the transfer matrix, while in the derivative-based basis it appears in an equivalent symplectic form.

	For the remainder of this paper, we shall confine ourselves to the case of piecewise-homogeneous multilayers. The ordered exponential then reduces to a matrix product of layer or interface factors,
	\begin{equation}
		\T_{mn} = \T_{m\,m+1}\T_{m+1\,m+2}\cdots \T_{n-1\,n},
	\end{equation}
	where $1\le m \le n \le N$, $N$ being the total number of layers. In the following section we derive the components of $\T$ in terms of  transmission and reflection amplitudes.
	\subsection{Transfer matrix in terms of scattering amplitudes}
	In the previous section we used the first-order state vector $\bPhi$, built from the wave function and its derivative, in order to emphasise the spatial-evolution viewpoint. In this section we pass to the four-component amplitude vector $\bPsi$ of right- and left-moving modes, which is more natural for the scattering interpretation and for the transfer-matrix block structure. Thus we define
	\begin{equation}
		\bPsi_n =
			\begin{pmatrix}
			\balp_n\\
			\bbet_n
			\end{pmatrix},
		\quad
		\balp_n,\bbet_n\in\mathbb C^2, \label{eq:spinor}
	\end{equation}
	where \(\balp_n=\left(\alpha^{\uparrow}, \alpha^{\downarrow}\right)^T\) and \(\bbet_n=\left(\beta^{\uparrow}, \beta^{\downarrow}\right)^T\) denote right- and left-moving spin amplitudes in a chosen spin basis. In the simple parabolic-band model used later for explicit examples, these amplitudes multiply plane waves with spin-dependent wave vectors \(k^\uparrow,k^\downarrow\). More general spin-orbit or texture effects modify the local mode structure, but not the block scattering form derived below. Then the matching conditions at the interface can be stated as follows
	\begin{equation}
	\begin{aligned}\begin{pmatrix}
			\balp_{j}\\
			\bbet_{j}
		\end{pmatrix} & =\T_{j\;j+1}\begin{pmatrix}
			\balp_{j+1}\\
			\bbet_{j+1}
		\end{pmatrix}.
	\end{aligned} \label{eq:t_matrix_def}
\end{equation}
	The form of $\T$ in terms of transmission and reflection amplitudes can be established by considering boundary conditions at the left and right lead, corresponding to waves of unit amplitude arriving from infinity. We thus obtain the following equations
	\begin{equation}
		\begin{aligned}\begin{pmatrix}
				\one\\
				\rmat
			\end{pmatrix} & =\T_{j\;j+1}\begin{pmatrix}
				\tmat'\\
				\zero
			\end{pmatrix},\\[10pt]
			\begin{pmatrix}
				\zero\\
				\tmat
			\end{pmatrix} & =\T_{j\;j+1}\begin{pmatrix}
				\rmat'\\
				\one
			\end{pmatrix}.
		\end{aligned} \label{eq:t_matrix_deriv}
	\end{equation}
	Solving (\ref{eq:t_matrix_deriv}) for the blocks of $\T_{j\;j+1}$ we obtain
	\begin{equation}
		\T_{j\;j+1}=\begin{pmatrix}
			{\tmat'}^{-1} & -{\tmat'}^{-1}\rmat'\\
			\rmat{\tmat'}^{-1} & \tmat-\rmat{\tmat'}^{-1}\rmat'
		\end{pmatrix}. \label{eq:tm_struct}
	\end{equation}
	Using \eqref{eq:tm_struct} we can now write the composition law
	\begin{equation*}
		\T_{mp} = \T_{mn}\T_{np}
	\end{equation*}
	for $m<n<p$.
	Performing multiplication on the right-hand side and comparing both sides block-wise we obtain
	\begin{subequations} \label{eq:transfer_mp}
		\begin{align}
			{\tmat'}_{mp} &= {\tmat'}_{np}(\one-\rmat'_{mn}\rmat_{np})^{-1}{\tmat'}_{mn}, \label{eq:t_prime_mp}\\
			\rmat'_{mp} &= \rmat'_{np}+{\tmat'}_{np}(\one - \rmat'_{mn}\rmat_{np})^{-1}\rmat'_{mn}\tmat_{np}, \label{eq:r_prime_mp}\\
			\tmat_{mp} &= \tmat_{mn}(\one-\rmat_{np}\rmat'_{mn})^{-1}\tmat_{np}, \label{eq:t_mp}\\
			\rmat_{mp} &= \rmat_{mn}+\tmat_{mn}\rmat_{np}(\one-\rmat'_{mn}\rmat_{np})^{-1}{\tmat'}_{mn},
			\label{eq:r_mp}
		\end{align}
	\end{subequations}
	which provide the recursive composition law for the scattering amplitudes of an arbitrary stack assembled from two sub-stacks.
	\subsection{Current-preserving bilinear form}
	The precise matrix form of the current-conservation constraint depends on the
	basis used for the wave amplitudes. In a raw right/left-moving amplitude basis,
	where the amplitudes are not flux-normalised, the current form contains explicit
	wave-vector factors,
	\[
	\Jmat_n =
	\begin{pmatrix}
		\kmat_n & \zero\\
		\zero & -\kmat_n
	\end{pmatrix},
	\]
	with \(\kmat_n=\mathrm{diag}(k_n^\uparrow,k_n^\downarrow)\). Thus a transfer
	matrix mapping amplitudes from layer \(n\) to layer \(m\) satisfies
	\begin{equation}
	\T_{mn}^\dagger \Jmat_m \T_{mn}=\Jmat_n. \label{eq:current-form}
	\end{equation}
	After passing to flux-normalised amplitudes
	\[
	\Lmat_n=
	\begin{pmatrix}
		\kmat_n^{1/2} & \zero\\
		\zero & \kmat_n^{1/2}
	\end{pmatrix},
	\qquad
	\tilde{\bPsi}_n=\Lmat_n\bPsi_n,
	\]
the current form becomes the fixed
	indefinite metric \(\eta=\mathrm{diag}(\one,-\one)\), and the corresponding transfer
	matrix satisfies 
	\begin{equation}
	\tilde{\T}_{mn}=\Lmat_m\T_{mn}\Lmat_n^{-1},
	\qquad
	\tilde{\T}_{mn}^{\dagger}\eta\tilde{\T}_{mn}=\eta.
	\label{eq:flux_normalised_t}
	\end{equation}
	The $\mathbf{\Omega}_{\Phi}$ encountered in the derivative-based first-order formulation is another representation of the same conserved current. If \(\bPhi_n=\Xmat_n\bPsi_n\),
	then the current forms are related by
	\[
	\Jmat_n = \Xmat_n^\dagger \Jmat_\Phi \Xmat_n,
	\]
	up to the conventional overall prefactor. Thus the derivative basis, raw
	amplitude basis, and flux-normalised amplitude basis represent the same
	conservation law with different matrices.	
	The transfer matrix therefore belongs to a noncompact Lie group preserving an indefinite current form.
	This makes available the standard decompositions of noncompact groups, in particular the Iwasawa decomposition.
	
\section{M\"obius action and Iwasawa interpretation}
	In this section we investigate the consequences of the group-theoretic constraint established by \eqref{eq:current-form}. We build up the discussion to the introduction of Weyl chamber variables, which will play an important role in characterising the action of a multilayer in later examples.
	\subsection{Action on reflection matrices}
Let
\[
\T =
\begin{pmatrix}
	\ba&\bb\\
	\bc&\bd
\end{pmatrix}.
\]
Consider a two-dimensional family of solutions, with right- and left-moving
spin amplitudes arranged as \(2\times2\) matrices \(\balp\) and \(\bbet\). In the chart where \(\balp\) is invertible, define
\begin{equation}
\zmat=\bbet\balp^{-1}. \label{eq:projective_z}
\end{equation}
Thus
\[
\begin{pmatrix}
	\balp\\
	\bbet
\end{pmatrix}
=
\begin{pmatrix}
	\one\\
	\zmat
\end{pmatrix}
\balp .
\]
Acting with $\T$ gives
\begin{equation}
\zmat\mapsto\zmat'
=
(\bc+\bd\zmat)(\ba+\bb\zmat)^{-1}. \label{eq:matrix_moebius}
\end{equation}
Thus the transfer matrix acts projectively on reflection matrices by a matrix
M\"obius transformation. Matrix-valued M\"obius transformations also arise naturally in related 
recursive Green's-function constructions, for example in the adlayering 
approach to surface Green's functions developed by Umerski~\cite{PhysRevB.55.5266}.
The mapping \eqref{eq:matrix_moebius} naturally generalises the picture of the action of a transfer matrix as a motion in the unit disc \cite{SANCHEZSOTO2012191} to the spin-resolved case.
\subsection{Iwasawa decomposition}	
The transfer matrix of a lossless layered system belongs, after a suitable choice of basis, to a noncompact matrix group. This is the basic reason why decompositions such as the Iwasawa and Cartan are useful. The Iwasawa decomposition separates a transfer matrix into
three qualitatively different actions
\[
\T = \Kmat \Amat \Nmat ,
\]
where \(\Kmat\) is compact, \(\Amat\) is Abelian noncompact, and \(\Nmat\) is
unipotent, with nilpotent Lie algebra \cite{helgason2001differential,KnappLieGroupsBeyondIntroduction}.

The compact factor \(\Kmat\) preserves the current metric without producing
radial amplification or attenuation. In a spin-resolved transport problem it
therefore represents rotations of the channel frame: spin precession, spin-basis
changes, and phase-like mixing of right- and left-moving modes.

The Abelian factor \(\Amat\) contains explicit noncompact scaling in the
Iwasawa coordinate system. In the simplest spin-filtering picture it acts like a
channel-dependent boost, increasing one amplitude component relative to another.
It is therefore naturally associated with spin-selective filtering and
attenuation.

The factor \(\Nmat\) is the least familiar part physically. In the
projective action on reflection matrices it produces shear-like, horocyclic
motion. In layered transport this factor is naturally associated with
order-dependent mixing, multiple-reflection shear, and left--right asymmetric
features of the projective scattering map. A non-trivial \(\Nmat\) is generic in
multilayer scattering and should not by itself be interpreted as a signature of
a particular microscopic interaction. Rather, it provides a natural place in the
decomposition for effects that are not pure rotations and not purely radial
filtering. Chiral, interfacial, or structurally asymmetric contributions may
therefore appear as symmetry-odd components of this shear factor, although their
identification requires comparing systems related by the appropriate symmetry
operation.

Thus the Iwasawa factors organise the transfer matrix into rotations, scaling,
and shear. The subsequent Weyl-chamber projection deliberately discards much of
the \(\Kmat\)- and \(\Nmat\)-factor information in order to retain only the
minimal radial filtering data.

In the numerical sections below we do not attempt to plot the full \(\Kmat\), \(\Amat\),
	and \(\Nmat\) data. Unlike the spinless case\cite{SANCHEZSOTO2012191}, the higher-dimensional spin-resolved picture does not, in general, admit a straightforward representation on the unit disc. We use the Iwasawa decomposition as an interpretive
	factorisation. In Appendix~\ref{app:single-interface-cartan} we provide an elementary analytic derivation of the Iwasawa factors based on wave-function matching in the parabolic model. The plotted invariants, on the other hand, are the lower-dimensional Cartan/Weyl radial variables. We now turn to this radial projection.
	\section{Weyl-chamber variables}
	
The Weyl-chamber variables used in what follows are radial invariants of the
transfer matrix. More precisely, they are associated with the Cartan projection
of the current-preserving transfer group. This distinction is important because
the Iwasawa decomposition \(g=kan\) and the Cartan decomposition
\(g=k_1a_ck_2\) both involve an Abelian noncompact subgroup conventionally
denoted by \(A\), but the corresponding \(A\)-components are not the same
function of \(g\)\cite{Duistermaat1984IwasawaProjection}. The Iwasawa decomposition is useful for interpreting
rotations, scaling, and shear, whereas the Cartan projection isolates the
bi-\(K\)-invariant noncompact data. It is this latter radial data that is
represented by a point in the Weyl chamber.
	In the bounded-domain realisation relevant to reflection matrices, these radial variables are obtained directly from the singular values of the reflection matrix.
	In practice we do not require an explicit numerical \(KAK\) decomposition. The M\"obius action introduced above acts naturally on a matrix-valued
	projective coordinate \(\zmat\) \eqref{eq:projective_z}. 
	For a current-conserving transfer matrix this action preserves the bounded
	matrix ball, so the singular values of \(\zmat\) satisfy
	\[
	0\leq\sigma_2\leq\sigma_1<1.
	\]
	The corresponding radial variables are
	\[
	\lambda_i=\operatorname{artanh}\sigma_i,
	\qquad i=1,2.
	\]
	The ordered pair
	\[
	\lambda_1\geq\lambda_2\geq0
	\]
	defines a point in the rank-two Weyl chamber.
	
	The physical reflection matrix is obtained by imposing the scattering boundary
	condition. For incidence from the left one has
	\[
	\begin{pmatrix}
		\balp\\
		\bbet
	\end{pmatrix}
	=
	\begin{pmatrix}
		\one\\
		\rmat
	\end{pmatrix},
	\]
	and hence the projective coordinate is simply $\zmat=\rmat$.
	Thus the Weyl coordinates used in the numerical examples are the radial
	coordinates of the physical reflection point in the matrix ball.
	
\section{Spin--Mueller--Jones representation}
The Weyl-chamber variables introduced above retain only the radial,
bi-\(K\)-invariant part of the transfer matrix. They therefore do not describe
how charge and spin current components are rotated or mixed. To keep track of
this observable-level information, it is useful to introduce a real
representation of the same transfer-matrix action on the space of Hermitian
direction--spin observables. This is the spin analogue of the Mueller--Jones
construction in polarisation optics.

Let
\[
\Gamma_{\mu\nu}=\tau_\mu\otimes\sigma_\nu,
\qquad
\mu,\nu=0,x,y,z,
\]
where \(\tau_\mu\) acts on the right/left-moving degree of freedom and
\(\sigma_\nu\) acts on spin. A coherent direction--spin density matrix may be
expanded as
\[
\rho
=
\frac14
\sum_{\mu,\nu}
\mathfrak J_{\mu\nu}\Gamma_{\mu\nu},
\qquad
\mathfrak J_{\mu\nu}
=
\operatorname{tr}(\rho\Gamma_{\mu\nu}).
\]
The transfer matrix acts by
\[
\rho\mapsto \rho'=\T\rho\T^\dagger .
\]
This induces a real \(16\times16\) matrix
\[
\mathfrak J'_{\mu\nu}
=
\sum_{\lambda,\rho}
\mathfrak M_{\mu\nu,\lambda\rho}(\T)
\mathfrak J_{\lambda\rho},
\]
where
\[
\mathfrak M_{\mu\nu,\lambda\rho}(\T)
=
\frac14
\operatorname{tr}
\left[
\Gamma_{\mu\nu}
\T
\Gamma_{\lambda\rho}
\T^\dagger
\right].
\]
is the Spin--Mueller--Jones tensor. Because matrix multiplication is associative, this construction satisfies
\[
\mathfrak M(\T_2\T_1)
=
\mathfrak M(\T_2)\mathfrak M(\T_1).
\]
Thus the Spin--Mueller--Jones tensor is a real observable-space representation
of the transfer-matrix group action.
The components with \(\mu=z\) correspond to right-minus-left flow. Thus \(\mathfrak J_{z0}\) is the charge-current component, while \(
\left(
\mathfrak J_{zx},
\mathfrak J_{zy},
\mathfrak J_{zz}
\right)
\)
are the spin current components. These are precisely the observables entering
the spin current imbalance used to define spin-transfer torque.

The ordinary Mueller--Jones construction has a special representation-theoretic
interpretation: the action of \(SL(2,\mathbb C)\) on \(2\times2\)
Hermitian matrices realises the Lorentz-vector representation
\((1/2,1/2)\). This is the origin of the familiar homomorphism
\(SL(2,\mathbb C)\to SO^+(1,3)\). In the present problem the transfer matrix
acts on \(\mathbb C^2_{\rm dir}\otimes\mathbb C^2_{\rm spin}\), so the
observable space is \(\mathrm{Herm}(4)\). If the transfer matrix factorises as
\(\T=\ba\otimes \bb\), the induced Spin--Mueller tensor factorises as
\[
\mathfrak M(\T)=\Mmat(\ba)\otimes \Mmat(\bb),
\]
and \(\mathfrak J_{\mu\nu}\) transforms as a rank-two Lorentz tensor under
\(SO^+(1,3)\times SO^+(1,3)\). A general spin-dependent
transfer matrix, however, mixes direction and spin and therefore belongs to the
larger \(4\times4\) congruence action. The Spin--Mueller--Jones tensor should thus be
viewed as an extension of the Mueller--Jones observable representation, not as
an object fully classified by the usual \(SL(2,\mathbb C)\) half-integer
weights.

The Spin--Mueller--Jones tensor is not used below as a replacement
for the Weyl-chamber variables. Instead, the two constructions are complementary:
the Weyl coordinates retain the radial filtering data, while the
Spin--Mueller--Jones tensor records the induced real action on charge and
spin current observables.
\section{Baker--Campbell--Hausdorff expansion and torque structures}
In Section~\ref{subsec:spatial-evolution} it was shown, in very general terms, how the ordered product of layer transfer matrices follows directly from the
spatially ordered exponential. For piecewise-constant layers one may write
\[
\T_j=\exp \Xmat_j,
\]
so that the total transfer matrix is
\[
\T=\T_N\cdots\T_1.
\]
For two factors,
\[
\T_2\T_1
=
\exp(\Xmat_2)\exp(\Xmat_1)
=
\exp(\Xmat_{\mathrm{eff}}),
\]
where the Baker--Campbell--Hausdorff expansion gives
\[
\Xmat_{\mathrm{eff}}
=
\Xmat_2+\Xmat_1
+\frac12[\Xmat_2,\Xmat_1]
+\frac1{12}[\Xmat_2,[\Xmat_2,\Xmat_1]]
+\cdots .
\]
The commutator terms are the Lie-algebraic counterpart of the noncommuting
reflection products that appear in the multiple-reflection expansion discussed
below.

To illustrate the spin structure, consider two spin-dependent generators of the
schematic form
\[
\Xmat_i=\gamma_i\,\vct m_i\cdot\bsig,
\]
where \(\vct m_i\) are magnetisation directions and \(\gamma_i\) are complex
coefficients encoding the strength and phase of the corresponding spin filter. Strictly speaking, the BCH generators act on the full
direction--spin space
\(\mathbb C^2_{\rm dir}\otimes\mathbb C^2_{\rm spin}\).
Thus the expressions below should be understood as the spin part of
generators of the form
\[
\Xmat_i=\gamma_i\bd_i\otimes(\vct m_i\cdot\boldsymbol\sigma),
\]
where \(\bd_i\) acts on the right/left-moving degree of freedom. When the
direction-space factors are identical or commute, the spin commutator reduces
to the familiar Pauli algebra. The corresponding
\(4\times4\) direction--spin calculation is given in
Appendix~\ref{app:bch-direction-spin}.
Using
\[
[
\vct a\cdot\bsig,
\vct b\cdot\bsig
]
=
2\ii(\vct a\times\vct b)\cdot\bsig,
\]
the leading commutator is
\[
[\Xmat_2,\Xmat_1]
=
2\ii\gamma_1\gamma_2
(\mathbf m_2\times\mathbf m_1)\cdot\bsig .
\]
For two in-plane magnetisations this term is out of plane. It therefore has the
spin structure associated with a field-like torque contribution.

The next nested commutator returns an in-plane structure. For example,
\[
[\Xmat_1,[\Xmat_2,\Xmat_1]]
=
-4\gamma_1^2\gamma_2\,
\mathbf m_1\times(\mathbf m_2\times\mathbf m_1)\cdot\bsig,
\]
up to the sign convention used for \(\Xmat_i\). This is the vector structure
associated with the damping-like torque component. Thus the BCH hierarchy
naturally alternates between the axial structure
\[
\mathbf m_1\times\mathbf m_2
\]
and the in-plane structure
\[
\mathbf m_1\times(\mathbf m_2\times\mathbf m_1),
\]
with higher nested commutators generating higher angular harmonics. The standard torque decomposition may therefore be written as
\[
\boldsymbol{\tau}
=
\tau_{\mathrm{DL}}\,
\mathbf m_1\times(\mathbf m_2\times\mathbf m_1)
+
\tau_{\mathrm{FL}}\,
\mathbf m_1\times\mathbf m_2 .
\]
The BCH expansion does not by itself determine the coefficients
\(\tau_{\mathrm{DL}}\) and \(\tau_{\mathrm{FL}}\). These coefficients must be
computed from the full scattering amplitudes or from the corresponding
spin current imbalance. Its usefulness here is instead structural: it explains
why noncollinear spin filters and spin--orbit terms generate precisely the
commutator directions that appear in the torque decomposition.

\subsection{Spin--orbit coupling as an additional noncommuting generator}

Spin--orbit coupling may be incorporated at the effective transfer-matrix level
as an additional spin-dependent factor
\[
\T_{\mathrm{SOC}}=\exp \Xmat_{\mathrm{SOC}},
\qquad
\Xmat_{\mathrm{SOC}}\sim \ii\,\vct a\cdot\bsig ,
\]
where \(\vct a\) is an effective spin--orbit axis. Its commutator with a
spin-filtering generator \(\Xmat_{\mathrm{FM}}\sim\gamma\,\mathbf m\cdot\bsig\) is
\[
[\Xmat_{\mathrm{SOC}},\Xmat_{\mathrm{FM}}]
\sim
[\ii\,\vct a\cdot\bsig,\gamma\,\mathbf m\cdot\bsig]
=
-2\gamma\,(\vct a\times\mathbf m)\cdot\bsig .
\]
Thus SOC generates an additional noncommuting spin direction. In a multilayer
this modifies both the compact spin-rotation sector and the noncompact filtering
sector after the full product is formed. This provides a natural algebraic
mechanism for enhanced field-like torque and for the curvature of Weyl-chamber
trajectories observed in the numerical examples below.

It is important to distinguish this commutator analysis from the Weyl-chamber
projection. The BCH expansion describes how the ordered product of physical
layer factors generates new Lie-algebra directions. The Weyl variables, by
contrast, are obtained only after the full transfer matrix has been formed and
projected to its Cartan radial data. Thus the BCH expansion explains the origin
of torque-relevant compact and noncompact components, while the Weyl trajectory
records the final radial filtering content of the resummed multilayer.

\section{Analytic spin-filter examples}

The Weyl-chamber construction is most transparent in simple spin-filtering
examples. In this section we first consider a single ferromagnetic layer between
identical nonmagnetic leads. We then consider two spin filters separated by a
normal spacer and show how the first noncommuting reflection product leads
naturally to the commutator structures introduced in the BCH discussion above.

\subsection{Single ferromagnetic layer}

Consider a single uniformly magnetised ferromagnetic layer of thickness \(L\)
between identical nonmagnetic leads. In the spin basis aligned with the
magnetisation, the reflection matrix is diagonal:
\[
\rmat_0(L)
=
\begin{pmatrix}
	\rho^\uparrow(L) & 0\\
	0 & \rho^\downarrow(L)
\end{pmatrix}.
\]
For a parabolic-band matching model, the scalar reflection amplitude in spin
channel \(s=\uparrow,\downarrow\) may be written, up to an irrelevant phase, as
\[
\rho^s(L)
=
\frac{
	\ii\left(k^2-(k_s)^2\right)\sin(k_sL)
}{
	2kk_s\cos(k_sL)
	-
	\ii\left(k^2+(k_s)^2\right)\sin(k_sL)
}.
\]
Here \(k\) is the wave vector in the nonmagnetic leads and \(k_s\) is the
spin-dependent wave vector inside the ferromagnet. If the magnetisation is rotated by an in-plane angle \(\theta\), then
\[
\rmat(\theta,L)
=
\smat^{-1}(\theta)\,
\rmat_0(L)\,
\smat(\theta),
\]
where
\[
\smat(\theta)
=
\exp\!\left(-\frac{\ii\theta}{2}\sigma_y\right).
\]
Since this is a unitary change of spin basis, the eigenvalues of
\(\rmat^\dagger\rmat\) are independent of \(\theta\):
\[
\operatorname{Spec}(\rmat^\dagger\rmat)
=
\left\{
|\rho^\uparrow(L)|^2,
|\rho^\downarrow(L)|^2
\right\}.
\]
Therefore the Weyl-chamber variables are
\[
\lambda_\uparrow(L)
=
\operatorname{artanh}|\rho^\uparrow(L)|,
\qquad
\lambda_\downarrow(L)
=
\operatorname{artanh}|\rho^\downarrow(L)|,
\]
ordered as
\[
\lambda_1=\max(\lambda_\uparrow,\lambda_\downarrow),
\qquad
\lambda_2=\min(\lambda_\uparrow,\lambda_\downarrow).
\]
This example separates the compact and noncompact information cleanly. The
magnetisation angle \(\theta\) changes the spin frame, hence the compact sector,
but does not change the Weyl-chamber point. By contrast, changing the layer thickness \(L\) or the Fermi energy level,
or the exchange splitting changes the scalar reflection amplitudes and therefore
moves the point in the Weyl chamber.
\subsection{Two spin filters and the first noncommuting reflection product}

The first genuinely spinful effect appears when two spin-dependent reflectors
are not diagonal in the same spin basis. Let the left spin filter, including the
layers up to region \(3\), have right-reflection matrix
\[
\rmat'_{13}(\theta)
=
\smat^{-1}(\theta)
\begin{pmatrix}
	\rho^\uparrow & 0\\
	0 & \rho^\downarrow
\end{pmatrix}
\smat(\theta),
\]
and let the right spin filter, consisting of the layers from region \(3\) to the
right lead, have left-reflection matrix
\[
\rmat_{35}
=
\begin{pmatrix}
	\chi^\uparrow & 0\\
	0 & \chi^\downarrow
\end{pmatrix}.
\]
Using the factorisation $\T_{15}=\T_{13}\T_{35}$ and \eqref{eq:transfer_mp} it can be shown that the total right-moving amplitude in the spacer region is given by
\[
\overrightarrow{\ba}_3
=
\left(\one-\rmat'_{13}(\theta)\rmat_{35}\right)^{-1}
\tmat'_{13}(\theta).
\]
Expanding the inverse gives
\[
\overrightarrow{\ba}_3
=
\tmat'_{13}
+
\rmat'_{13}\rmat_{35}\tmat'_{13}
+
(\rmat'_{13}\rmat_{35})^2\tmat'_{13}
+\cdots .
\]
The first correction term is the first contribution involving reflections from
both spin filters.

The noncommutativity of the two spin filters is measured by
\[
[\rmat'_{13}(\theta),\rmat_{35}].
\]
Writing \(c=\cos(\theta/2)\) and \(s=\sin(\theta/2)\), one obtains
\[
\rmat'_{13}(\theta)
=
\begin{pmatrix}
	\rho^\uparrow c^2+\rho^\downarrow s^2
	&
	(\rho^\downarrow-\rho^\uparrow)cs\\
	(\rho^\downarrow-\rho^\uparrow)cs
	&
	\rho^\uparrow s^2+\rho^\downarrow c^2
\end{pmatrix}.
\]
Therefore
\[
[\rmat'_{13}(\theta),\rmat_{35}]
=
(\rho^\downarrow-\rho^\uparrow)(\chi^\downarrow-\chi^\uparrow)
cs
\begin{pmatrix}
	0&1\\
	-1&0
\end{pmatrix}.
\]
Equivalently,
\[
[\rmat'_{13}(\theta),\rmat_{35}]
=
\frac12
(\rho^\downarrow-\rho^\uparrow)
(\chi^\downarrow-\chi^\uparrow)
\sin\theta\,
\ii\sigma_y .
\]
Thus the commutator vanishes if the magnets are collinear, if either reflector
is spin independent, or if the two spin filters are diagonal in the same spin
basis. For in-plane magnetisations, the commutator points in the out-of-plane
spin direction. This is the amplitude-level origin of the out-of-plane spin
current generated by multiple reflections between noncollinear filters.
	\section{Computational examples}
	In the following examples the Weyl coordinates are
	computed from the reflection block via the bounded-domain prescription \(\lambda_i=\operatorname{artanh}\sigma_i(\rmat)\). The model structures are
	kept deliberately minimal so that the geometric content of each plot is visible: a single spin filter, a noncollinear spin valve, and a controlled spin-orbit mixing factor in the normal spacer.
	\subsection{Single FM layer between identical NM leads}
\begin{figure}[htbp]
		\centering
		\includegraphics[width=\linewidth]{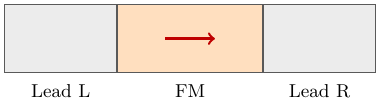}
		\caption{Model geometry for the single-FM example.  Identical
		nonmagnetic leads are attached to a single ferromagnetic layer whose
		thickness is swept.}
		\label{fig:single-fm-stack}
	\end{figure}

\begin{figure}[t]
		\centering
		\includegraphics[width=\linewidth]{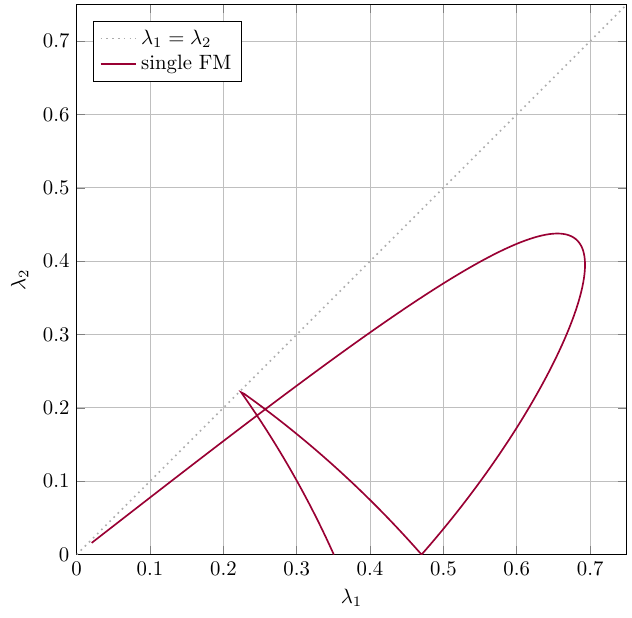}
		\caption{Weyl-chamber trajectory for the single-FM thickness sweep.  The
		diagonal \(\lambda_1=\lambda_2\) corresponds to spin-independent radial
		filtering.}
		\label{fig:single-fm-chamber}
	\end{figure}

	The first test is a single uniformly magnetised ferromagnetic layer between
	identical nonmagnetic leads, depicted in Figure~\ref{fig:single-fm-stack}.
	For the single-FM calculations we set \(k_\parallel=0\) and Fermi energy
	\(E=0.30\). The lead potential is \(V_{\rm lead}=0\).  The ferromagnetic
	layer has \(V_{\rm FM}=0.2\), exchange splitting \(\Delta_{\rm FM}=0.05\),
	and magnetisation angle \(\theta_{\rm FM}=0\).
	\begin{figure}[t]
		\centering
		\includegraphics[width=\linewidth]{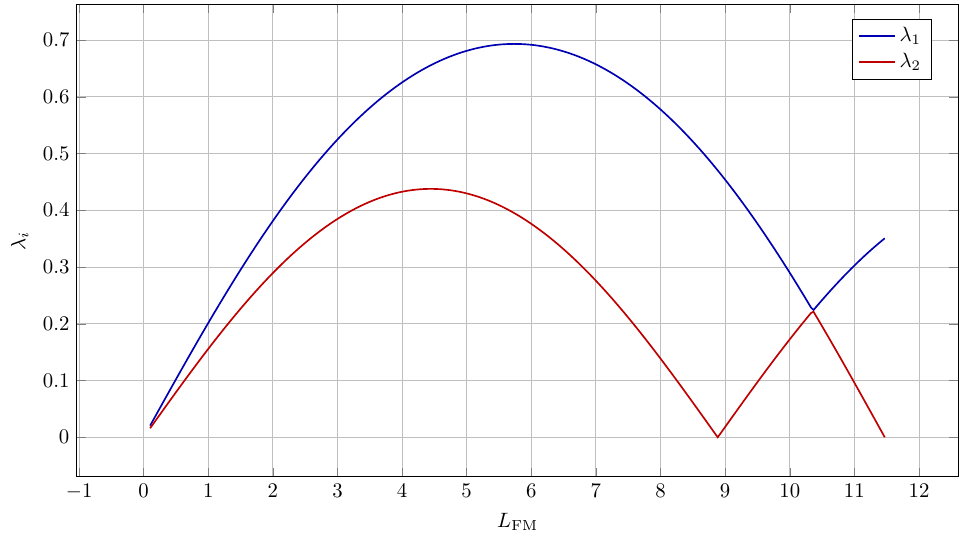}
		\caption{Single-FM Weyl coordinates as a function of ferromagnetic
		thickness.}
		\label{fig:single-fm-thickness}
	\end{figure}
	We plot the Weyl coordinates as functions of the FM layer thickness in Figure~\ref{fig:single-fm-thickness}. The same data may be represented as a path in the positive Weyl chamber, Figure~\ref{fig:single-fm-chamber}. The
	diagonal \(\lambda_1=\lambda_2\) marks spin-independent radial filtering, so
	departure from this diagonal gives a compact visual measure of spin-selective
	filtering.
	\subsection{FM/NM/FM trilayer without spin--orbit coupling}
\begin{figure*}[tp]
		\centering
		\includegraphics[width=\textwidth]{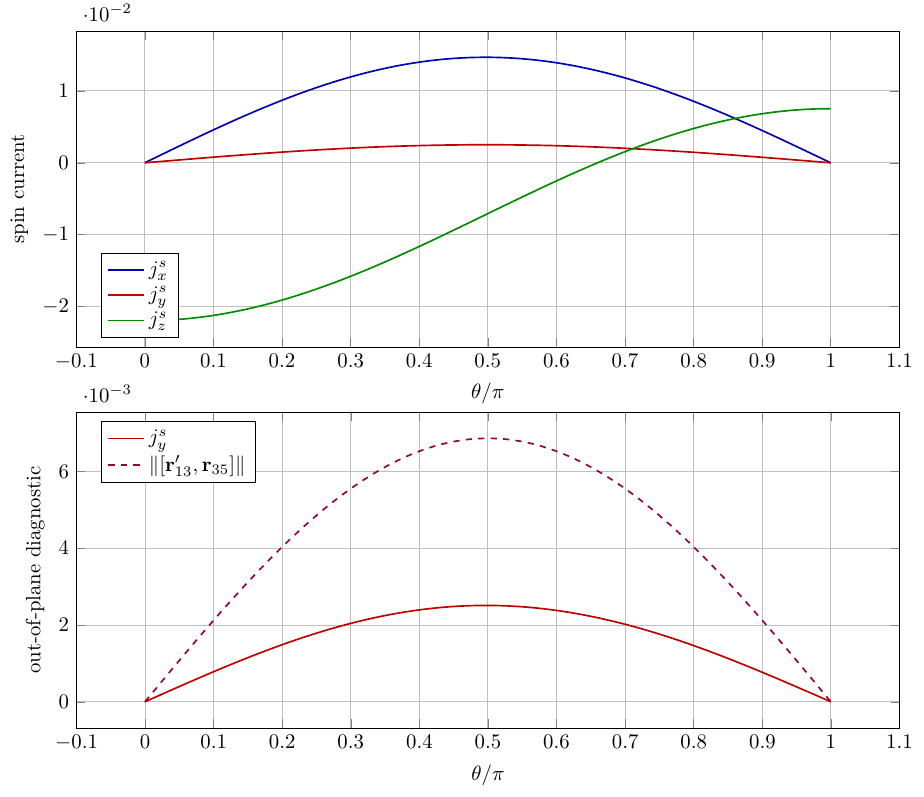}
		\caption{Spin-current components in the spacer of the FM/NM/FM stack as
		the analyser angle is swept. The lower panel compares the out-of-plane
		spin current component with the noncommutativity diagnostic
		\(\|[\rmat'_{13},\rmat_{35}]\|\).}
		\label{fig:spin-valve-current-commutator}
	\end{figure*}

\begin{figure}[t]
		\centering
		\includegraphics[width=\linewidth]{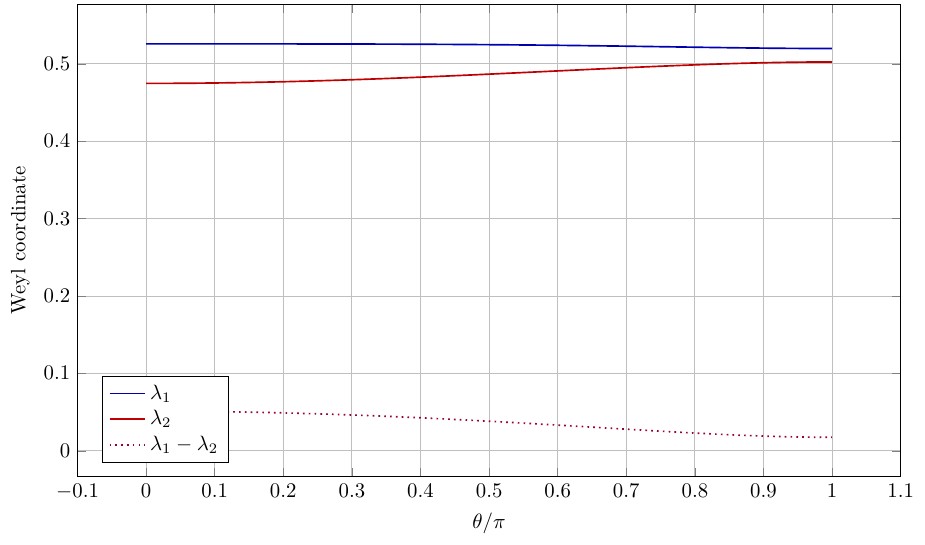}
		\caption{Weyl coordinates for the FM/NM/FM stack as a function of analyser
		angle. The splitting \(\lambda_1-\lambda_2\) tracks the anisotropic part
		of the radial filtering response.}
		\label{fig:spin-valve-weyl-angle}
	\end{figure}

	The next example is a noncollinear FM/NM/FM stack.  We sweep the analyser
	magnetisation angle while monitoring spin current components in the normal
	spacer. The lower panel in Figure~\ref{fig:spin-valve-current-commutator}
	compares the out-of-plane component \(j_y^s\) with the spacer reflection commutator norm
	\[
		\left\|[\rmat'_{13},\rmat_{35}]\right\|,
	\]
	where \(\rmat'_{13}\) is the right-incidence reflection block of the left sub-stack and \(\rmat_{35}\) is the left-incidence reflection block of the right sub-stack. Both vanish in the collinear limits and peak in the noncollinear regime.
	\begin{figure}[htbp]
		\centering
		\includegraphics[width=\linewidth]{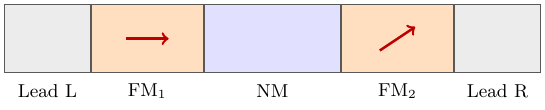}
		\caption{Model geometry for the FM/NM/FM spin-valve example. The first
		ferromagnet is used as a fixed polariser, while the second acts as an
		analyser whose angle or spacer separation is swept.}
		\label{fig:spin-valve-stack}
	\end{figure}
	For the FM/NM/FM calculations we set \(k_\parallel=0\) and
	\(E=0.35\). The leads have \(V_{\rm lead}=0\).  The left
	ferromagnet has \(V_1=0.2\), \(\Delta_1=0.03\), \(L_1=3\), and
	\(\theta_1=0\). The normal spacer has \(V_{\rm NM}=0\), with
	\(L_{\rm NM}=12\) for the analyser-angle sweep, Figure~\ref{fig:spin-valve-weyl-angle}. The right ferromagnet has
	\(V_2=0.2\), \(\Delta_2=0.04\), \(L_2=4\).
	The corresponding Weyl coordinates show how the same noncollinear composition changes the radial, noncompact filtering data. We also plot
	\(\lambda_1-\lambda_2\), which is a scalar measure of the spin-selective
	part of the radial projection.

\begin{figure}[t]
		\centering
		\includegraphics[width=\linewidth]{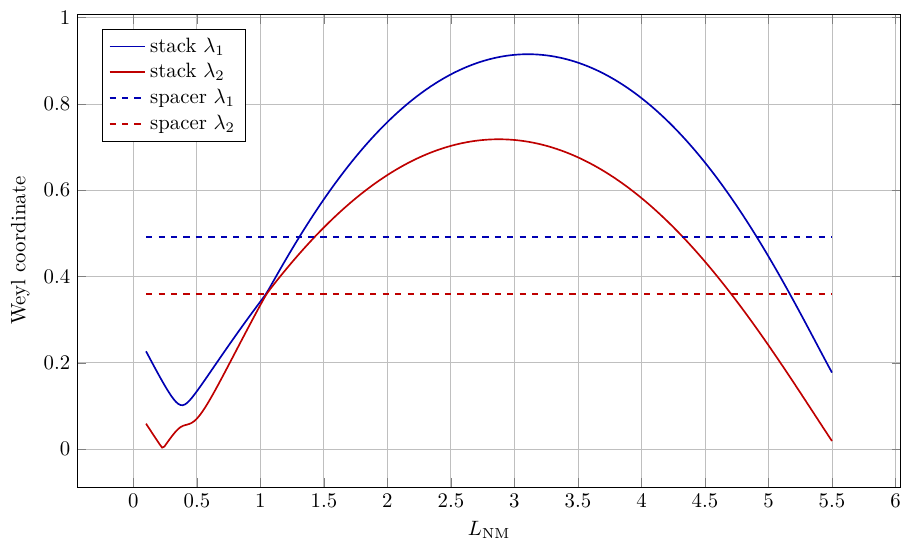}
		\caption{FM/NM/FM Weyl coordinates as a function of normal-spacer
		thickness. The dashed spacer-local reference is nearly stationary, while
		the full-stack Weyl coordinates oscillate with the spacer propagation
		phase.}
		\label{fig:spin-valve-spacer-thickness}
	\end{figure}

\begin{figure}[t]
		\centering
		\includegraphics[width=\linewidth]{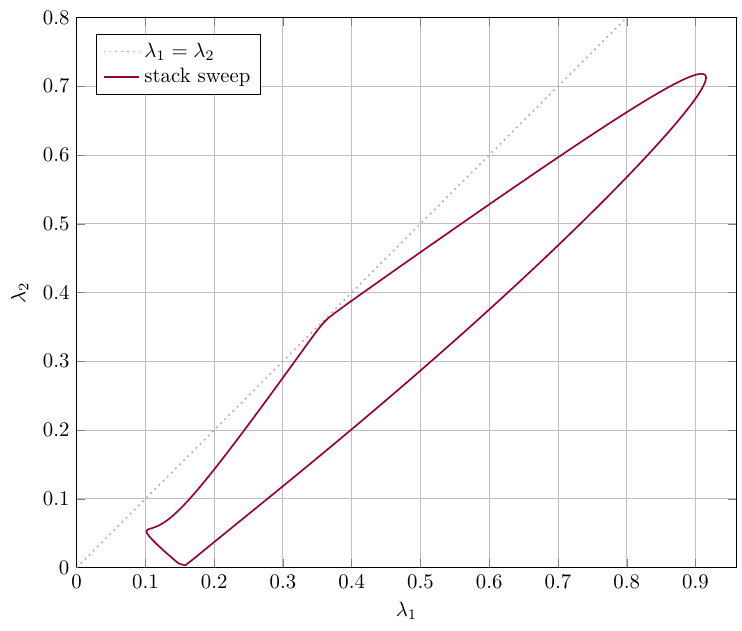}
	\caption{Weyl-chamber trajectory for the FM/NM/FM spacer-thickness sweep.}
		\label{fig:spin-valve-spacer-chamber}
	\end{figure}

	A complementary sweep varies the normal-spacer thickness while keeping the
	two ferromagnets fixed. This exposes the radial effect of the propagation
	phase accumulated between the two magnetic reflectors.  In Figure~\ref{fig:spin-valve-spacer-thickness}
	the solid curves show the Weyl coordinates of the full FM/NM/FM stack, while
	the dashed curves show the spacer-local right-reflection block as a reference.
	The same sweep gives a Weyl-chamber trajectory analogous to the single-FM
	thickness trajectory in Figure~\ref{fig:spin-valve-spacer-chamber}, but now the path is generated by the noncollinear
	composition of two magnetic reflectors separated by a normal spacer.
\subsection{FM/NM/FM trilayer with a controlled spin--orbit mixing factor}

\begin{figure*}[tp]
	\centering
	\includegraphics[width=\textwidth]{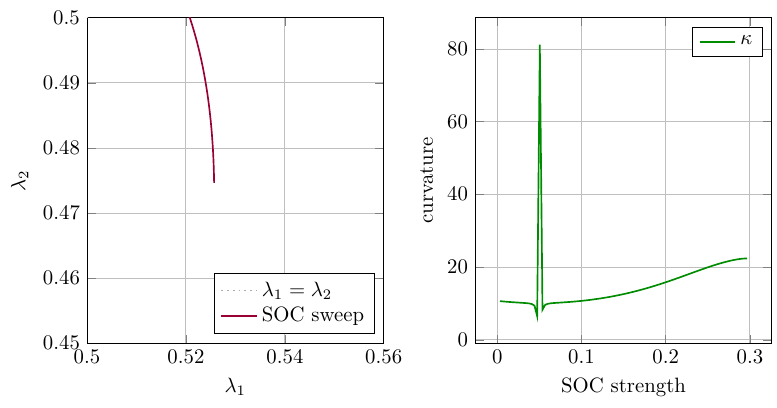}
	\caption{Controlled-SOC Weyl-chamber trajectory and curvature diagnostic for the FM/NM/FM stack.}
	\label{fig:soc-weyl-curvature}
\end{figure*}

Finally, we add a controlled spin--orbit mixing factor in the normal spacer in Figure~\ref{fig:soc-weyl-curvature}.
This is not yet a full layer-resolved SOC transfer model. Rather, it is a
minimal model insertion used to isolate the geometric effect of adding an
additional noncommuting spin direction. In the numerical implementation the
microscopic transfer matrix still ignores the stored layer SOC parameter; the
SOC sweep is implemented by inserting a compact spin-mixing factor in the
spacer transfer product.

Let the spacer be layer \(s\). We write the total transfer matrix as
\begin{equation}
	\T_{\rm tot}(\alpha_{\rm so})
	=
	\T_{L\to s}\,
	\Rmat_{\rm so}(\varphi_{\rm so})\,
	\T_{s\to R},
	\label{eq:controlled_soc_transfer}
\end{equation}
where
\begin{equation*}
	\varphi_{\rm so}
	=
	\alpha_{\rm so}L_s .
\end{equation*}
Here \(L_s\) is the spacer thickness and \(\alpha_{\rm so}\) is a controlled
spin-mixing strength per unit spacer length. In the example below the spacer
SOC vector is stored as
\begin{equation*}
	\vct A_s=(0,0,\alpha_{\rm so}),
\end{equation*}
so that
\begin{equation*}
	\alpha_{\rm so}=|\vct A_s|.
\end{equation*}
The inserted factor is block diagonal in the right/left-moving degree of
freedom,
\begin{equation}
	\Rmat_{\rm so}(\varphi_{\rm so})
	=
	\begin{pmatrix}
		\smat_{\rm so}(\varphi_{\rm so}) & \zero\\
		\zero & \smat_{\rm so}(\varphi_{\rm so})
	\end{pmatrix},
	\label{eq:soc_block_rotation}
\end{equation}
with
\begin{equation*}
	\smat_{\rm so}(\varphi_{\rm so})
	=
	\exp\!\left(
	-\frac{\ii}{2}
	\varphi_{\rm so}\,
	\hat{\vct n}_{\rm so}\cdot\bsig
	\right).
\end{equation*}
The unit vector \(\hat{\vct n}_{\rm so}\) specifies the spin-rotation axis used
in this controlled model. Thus \(\alpha_{\rm so}\) should not be interpreted as
a microscopic Rashba or Dresselhaus coupling.

For each value of \(\alpha_{\rm so}\), we form the scattering amplitudes from
\(\T_{\rm tot}(\alpha_{\rm so})\). The Weyl-chamber coordinates are obtained
from the reflection matrix,
\begin{equation*}
	\rmat^\dagger(\alpha_{\rm so})\rmat(\alpha_{\rm so})u_i
	=
	\sigma_i^2(\alpha_{\rm so})u_i,
	\qquad
	1>\sigma_1\geq\sigma_2\geq0,
\end{equation*}
by
\begin{equation}
	\lambda_i(\alpha_{\rm so})
	=
	\operatorname{artanh}\sigma_i(\alpha_{\rm so}).
	\label{eq:soc_weyl_coordinates}
\end{equation}
The SOC sweep therefore defines a parametrised curve
\begin{equation*}
	\vct\lambda(\alpha_{\rm so})
	=
	\bigl(
	\lambda_1(\alpha_{\rm so}),
	\lambda_2(\alpha_{\rm so})
	\bigr)
\end{equation*}
inside the Weyl chamber.

We also use the plane-curve curvature as a local diagnostic of bending. For a
smooth curve \(\vct\lambda(\alpha_{\rm so})\), this is
\begin{equation}
	\kappa(\alpha_{\rm so})
	=
	\frac{
		\left|
		\dot{\lambda}_1\ddot{\lambda}_2
		-
		\dot{\lambda}_2\ddot{\lambda}_1
		\right|
	}{
		\left(
		\dot{\lambda}_1^2+\dot{\lambda}_2^2
		\right)^{3/2}
	},
	\label{eq:weyl_curvature}
\end{equation}
where dots denote derivatives with respect to \(\alpha_{\rm so}\). This
curvature is used only as a geometric diagnostic of the parametrised Weyl
trajectory. In particular, large values can occur when the curve nearly stalls
in the chosen parameter, because the denominator in
\eqref{eq:weyl_curvature} depends on the cube of the speed.
	
	\section{Outlook}
	Here we outline some possible directions for further development and application of the formalism discussed in this paper.
	\subsection{Extension to orbital transport}
		For a \(p\)-orbital extension, the internal wave function belongs to
		\(\mathbb C^3_{\rm orb}\) rather than \(\mathbb C^2_{\rm spin}\). The transfer
		state is therefore
		\[
		\Psi_{\rm orb}
		=
		\begin{pmatrix}
			\boldsymbol\alpha\\
			\boldsymbol\beta
		\end{pmatrix}
		\in
		\mathbb C^2_{\rm dir}\otimes\mathbb C^3_{\rm orb}.
		\]
		For a fixed transverse momentum \(\mathbf k_\parallel\), one solves the local
		orbital mode problem
		\[
		H_n(k_x,\mathbf k_\parallel)u_{n,a}
		=
		E u_{n,a}
		\]
		to obtain right- and left-moving orbital eigenmodes. In each layer,
		\[
		\psi_n(x)
		=
		\sum_{a=1}^{3}
		\alpha_{n,a}u_{n,a}^{+}e^{ik_{x,n,a}^{+}x}
		+
		\sum_{a=1}^{3}
		\beta_{n,a}u_{n,a}^{-}e^{-ik_{x,n,a}^{-}x}.
		\]
		The corresponding \(6\times6\) transfer matrix is constructed by matching the
		wave function and the appropriate velocity/current across interfaces. After
		flux normalisation it preserves
		\[
		\eta_3=
		\begin{pmatrix}
			\one_3&\zero\\
			\zero&-\one_3
		\end{pmatrix},
		\]
		so the natural transfer group is of \(U(3,3)\)-type. The Weyl chamber is
		therefore rank three, with coordinates obtained from the singular values of the
		\(3\times3\) orbital reflection matrix,
		\[
		\lambda_i=\operatorname{artanh}\sigma_i(\rmat),
		\qquad
		i=1,2,3.
		\]
	The bilayer model\cite{sun2026interfacialorbitaltransmissionconversion}
	provides a concrete realisation of this structure, with the right layer
	described by a \(p\)-orbital Hamiltonian containing a crystal-field term
	\(r(\mathbf L\cdot\mathbf k)^2\), leading naturally to orbital dipole--quadrupole
	conversion and mechanical torque.
	\subsection{Extension to tight-binding models}
	In a realistic tight-binding calculation the number of propagating modes may be
	large, and the current-preserving transfer group is correspondingly enlarged
	from \(U(2,2)\) to \(U(N,N)\). The Weyl chamber is then \(N\)-dimensional.
	Although this is no longer directly visualisable, the same radial data can be
	extracted from the singular values of the reflection matrix. The resulting Weyl spectrum refines the usual conductance eigenvalue
	description: \(T_i=\operatorname{sech}^2\lambda_i\), so conductance is a scalar
	functional of the Weyl spectrum. For realistic multilayers one may reduce this
	data by symmetry sectors, transverse momentum, spin or orbital projections, or
	by retaining only the dominant transmission eigenchannels.
	\section{Conclusion}
	We have discussed coherent spin transport in magnetic multilayers using the
	geometric language of transfer matrices. Drawing on standard ideas from
	noncompact Lie groups, we described the action of a multilayer in terms of its
	projective Möbius action, Iwasawa interpretation, and Cartan/Weyl radial
	projection. The Weyl-chamber variables provide a compact description of the
	noncompact spin-filtering data, while the Spin--Mueller--Jones tensor gives a
	complementary observable-space representation acting on charge and spin current
	components. The examples show how this viewpoint separates compact spin-frame
	rotations, radial filtering, multiple-reflection effects, and controlled
	spin--orbit mixing. Possible extensions include orbital transport and
	higher-dimensional tight-binding mode spaces.
\appendix

\section{Single-interface transfer matrix in flux-normalised form}
\label{app:single-interface-cartan}
We provide a minimal wave-function matching derivation for a single
spin-dependent interface and show how the resulting transfer matrix acquires
a simple compact/noncompact form after flux normalisation.

Consider an interface at \(x=0\) between a nonmagnetic lead \(L\) and a
ferromagnetic layer \(F\). In the lead the spin-resolved wave-vector matrix is
proportional to the identity,
\[
\mathbf k_L = k\,\one ,
\]
while in the local spin basis of the ferromagnet
\[
\mathbf k_F=
\diag(k_+,k_-).
\]
We assume here that \(k,k_+,k_-\) are real and positive, so that all channels
are propagating. Evanescent channels require the corresponding current form to
be treated separately.

In each homogeneous region we write the two-component spin wave function as a
sum of right- and left-moving amplitudes,
\[
\vct\psi_n(x)
=
\balp_n e^{\ii \mathbf k_n x}
+
\bbet_n e^{-\ii \mathbf k_n x},
\]
where
\[
\balp_n,\bbet_n\in\mathbb C^2.
\]
Define the four-component amplitude vector
\[
\boldsymbol\Psi_n
=
\begin{pmatrix}
	\balp_n\\
	\bbet_n
\end{pmatrix}.
\]
At the interface, continuity of the wave function and its derivative gives
\[
\vct\psi_L(0)=\vct\psi_F(0),
\qquad
\partial_x\vct\psi_L(0)=\partial_x\vct\psi_F(0).
\]
These two equations may be written compactly as
\[
\mathbf X(\mathbf k_L)\boldsymbol\Psi_L
=
\mathbf X(\mathbf k_F)\boldsymbol\Psi_F,
\]
where
\[
\mathbf X(\mathbf k)
=
\begin{pmatrix}
	\one & \one\\
	\ii\mathbf k & -\ii\mathbf k
\end{pmatrix}.
\]
Thus the raw amplitude-basis transfer matrix is
\[
\boldsymbol\Psi_L
=
\mathbf T_{LF}\boldsymbol\Psi_F,
\qquad
\mathbf T_{LF}
=
\mathbf X^{-1}(\mathbf k_L)\mathbf X(\mathbf k_F).
\]

If the magnetisation of the ferromagnet is rotated relative to the spin basis
of the lead, we insert the spin-frame rotation
\[
\mathbf S(-\theta)
=
\begin{pmatrix}
	\mathbf s(-\theta)&\zero\\
	\zero&\mathbf s(-\theta)
\end{pmatrix},
\]
where \(\mathbf s(\theta)\) is the corresponding \(2\times2\) spin rotation.
Since the lead is nonmagnetic, \(\mathbf k_L=k\one\), this rotation commutes
through the lead matching matrix. The rotated interface matrix may therefore be
written as
\[
\mathbf T_{LF}(\theta)
=
\mathbf S(-\theta)
\mathbf X^{-1}(\mathbf k_L)\mathbf X(\mathbf k_F).
\]

The matrix \(\mathbf T_{LF}\) is not yet in a basis with a fixed current metric.
The flux carried by a right-moving channel is proportional to its wave vector,
so we introduce flux-normalised amplitudes
\[
\widetilde{\boldsymbol\Psi}_n
=
\mathbf L_n\boldsymbol\Psi_n,
\qquad
\mathbf L_n
=
\begin{pmatrix}
	\mathbf k_n^{1/2}&\zero\\
	\zero&\mathbf k_n^{1/2}
\end{pmatrix}.
\]
The flux-normalised transfer matrix is
\[
\widetilde{\mathbf T}_{LF}(\theta)
=
\mathbf L_L
\mathbf T_{LF}(\theta)
\mathbf L_F^{-1}.
\]
In this basis the current form is independent of the layer, and, for real propagating wave vectors,
\[
\widetilde{\mathbf T}_{LF}^{\dagger}
\eta
\widetilde{\mathbf T}_{LF}
=
\eta .
\]
A direct calculation gives
\[
\widetilde{\mathbf T}_{LF}(\theta)
=
\mathbf S(-\theta)\mathbf A_{LF},
\]
where
\[
\mathbf A_{LF}
=
\begin{pmatrix}
	\cosh\boldsymbol\chi & -\sinh\boldsymbol\chi\\
	-\sinh\boldsymbol\chi & \cosh\boldsymbol\chi
\end{pmatrix},
\]
with
\begin{equation}
\boldsymbol\chi
=
\frac12
\log\left(\frac{\mathbf k_F}{k}\right)
=
\diag(\chi_+,\chi_-),
\quad
\chi_\pm=
\frac12\log\left(\frac{k_\pm}{k}\right). \label{eq:chi}
\end{equation}
Equivalently,
\[
\mathbf A_{LF}
=
\exp\left[
-\tau_x\otimes\boldsymbol\chi
\right],
\]
up to the sign convention chosen for the direction-space Pauli matrix
\(\tau_x\).

Thus the elementary flux-normalised interface matrix has the form
\[
\widetilde{\mathbf T}_{LF}(\theta)
=
\Kmat(\theta)\Amat,
\]
with
\[
\Kmat(\theta)=\mathbf S(-\theta),
\quad
\Amat=\mathbf A_{LF}.
\]
The compact factor \(\Kmat(\theta)\) changes the spin frame, while the noncompact
factor \(\Amat\) is a spin-dependent boost determined by the mismatch between the
lead wave vector \(k\) and the two ferromagnetic wave vectors \(k_\pm\).

The boost appears in hyperbolic form because we are using the right/left-moving
basis, in which the fixed current metric is \(\eta=\diag(\one,-\one)\). If one
changes to the direction-space basis that diagonalises \(\tau_x\),
\[
\mathbf C
=
\frac1{\sqrt2}
\begin{pmatrix}
	\one&\one\\
	\one&-\one
\end{pmatrix},
\]
then
\[
\mathbf C^{-1}
\mathbf A_{LF}
\mathbf C
=
\begin{pmatrix}
	\exp(-\boldsymbol\chi)&\zero\\
	\zero&\exp(\boldsymbol\chi)
\end{pmatrix}.
\]
In this transformed basis the current metric is transformed as well; the
diagonal expression is simply the same noncompact boost written in a different
direction-space representation.

For a single perfect interface of this type there is no nilpotent shear factor. The flux-normalised matrix is just a noncompact spin-dependent boost followed by a compact spin rotation.
\subsection*{Cartan projection of the single-interface matrix}

The same example also makes the Cartan projection explicit. In the
flux-normalised basis the interface matrix has the form
\[
\widetilde{\T}_{LF}(\theta)
=
K_1 A_c K_2,
\quad
K_1=\mathbf S(-\theta),
\quad
A_c=\mathbf A_{LF},
\quad
K_2=\one .
\]
The corresponding Cartan radial variables are
\[
|\chi_+|,
\quad
|\chi_-|,
\]
ordered into the positive Weyl chamber as
\[
\lambda_1=\max(|\chi_+|,|\chi_-|),
\quad
\lambda_2=\min(|\chi_+|,|\chi_-|).
\]

This agrees with the reflection-matrix definition of the Weyl coordinates. For
a single spin channel the reflection amplitude of a flux-normalised interface is
\[
r_\pm
=
\frac{k-k_\pm}{k+k_\pm}.
\]
Using \eqref{eq:chi} one finds
\[
r_\pm
=
-\tanh\chi_\pm .
\]
Therefore
\[
\operatorname{artanh}|r_\pm|
=
|\chi_\pm|.
\]
Thus the Cartan coordinates obtained from the noncompact interface boost are
the same as the Weyl coordinates obtained from the singular values of the
reflection matrix.

This equality is special to the elementary single-interface example. It occurs
because the flux-normalised matrix is already a compact factor multiplied by an
element of the Cartan subgroup. For a general multilayer, the Iwasawa
\(A\)-coordinate and the Cartan radial coordinate are different functions of
the full transfer matrix. The numerical Weyl coordinates used in the main text
are the Cartan radial variables, extracted operationally from the singular
values of the reflection matrix.
\section{Reciprocity transpose relations in the raw amplitude basis}
\label{app:reciprocity-transpose}
Here we briefly comment on the additional weighted bilinear symplectic relation satisfied by $\T$. It relies on the assumption that the spin-dependent scattering problem belongs to the reciprocal real-symmetric class, as in the coplanar exchange model with no spin--orbit coupling or $\sigma_y$-type terms. It can be shown\cite{fadeev2019application}, under essentially the same assumptions that the following relations hold
\begin{equation}
\rmat_{nm}^T=\rmat_{nm},
\qquad
{\rmat'_{nm}}^T=\rmat'_{nm},
\qquad
{\tmat'_{nm}}^T
=
\kmat_n\kmat_m^{-1}\tmat_{nm}.
\label{eq:app-t-reciprocity-general}
\end{equation}
Taking the transpose of \eqref{eq:tm_struct} and using \eqref{eq:app-t-reciprocity-general} we obtain
\begin{equation}
\T_{nm}^T\mathbf{\Omega}_n\T_{nm}
=
\mathbf{\Omega}_m,
\label{eq:app-weighted-symplectic}
\end{equation}
where
\[
\mathbf{\Omega}_n
=
\begin{pmatrix}
    \zero&\kmat_n\\
	-\kmat_n&\zero
\end{pmatrix}.
\]
If the amplitudes are flux-normalised, or if the asymptotic layers have the
same wave-vector normalisation, this reduces to the standard symplectic form
\[
\widetilde{\T}_{nm}^T\mathbf{\Omega}\widetilde{\T}_{nm}
=
\mathbf{\Omega},
\qquad
\mathbf{\Omega}=
\begin{pmatrix}
	\zero&\one\\
	-\one&\zero
\end{pmatrix}.
\]

Equation~\eqref{eq:app-weighted-symplectic} is a bilinear transpose relation
expressing reciprocity. It should be distinguished from the Hermitian
current-conservation condition \eqref{eq:current-form}, or from its
flux-normalised form \eqref{eq:flux_normalised_t}. The latter follows from
flux conservation, whereas the former requires the additional reciprocal
real-symmetric structure of the scattering problem.

The transpose relations need not hold in the presence of spin--orbit coupling,
noncoplanar magnetic textures involving \(\sigma_y\), vector-potential terms,
or other nonreciprocal/chiral interactions. In such cases the pseudo-unitary
current-conservation structure remains the appropriate general constraint,
whereas the symplectic transpose constraint is no longer guaranteed.

\section{Direction--spin form of the BCH commutators}
\label{app:bch-direction-spin}

The BCH discussion in the main text suppresses the right/left-moving degree of
freedom in order to display the spin algebra. Here we record the corresponding
direction--spin form. The transfer matrices act on
\(
\mathbb C^2_{\rm dir}\otimes\mathbb C^2_{\rm spin},
\)
so the BCH generators are \(4\times4\) matrices. A schematic spin-dependent
generator should therefore be understood as
\[
\Xmat_i
=
\gamma_i\,\bd_i\otimes \smat_i,
\qquad
\smat_i=\vct m_i\cdot\bsig,
\]
where \(\bd_i\) acts on the right/left-moving degree of freedom and \(\smat_i\) acts
on spin.

For two such generators,
\[
\Xmat_1=\gamma_1\bd_1\otimes \smat_1,
\qquad
\Xmat_2=\gamma_2\bd_2\otimes \smat_2,
\]
the commutator is
\[
[\Xmat_2,\Xmat_1]
=
\gamma_1\gamma_2
\left(
\bd_2\bd_1\otimes \smat_2\smat_1
-
\bd_1\bd_2\otimes \smat_1\smat_2
\right).
\]
Equivalently,
\[
[\Xmat_2,\Xmat_1]
=
\frac{\gamma_1\gamma_2}{2}
\left(
\{\bd_2,\bd_1\}\otimes[\smat_2,\smat_1]
+
[\bd_2,\bd_1]\otimes\{\smat_2,\smat_1\}
\right).
\]
Using the Pauli identities
\[
[\smat_2,\smat_1]
=
2\ii
(\vct m_2\times\vct m_1)\cdot\bsig,
\]
and
\[
\{\smat_2,\smat_1\}
=
2(\vct m_2\cdot\vct m_1)\one,
\]
one obtains
\begin{equation}
\begin{aligned}
[\Xmat_2,\Xmat_1]
&=
\ii\gamma_1\gamma_2
\{\bd_2,\bd_1\}\otimes
\left[
(\vct m_2\times\vct m_1)\cdot\bsig
\right]\\
&+
\gamma_1\gamma_2
(\vct m_2\cdot\vct m_1)
[\bd_2,\bd_1]\otimes\one .
\label{eq:app-full-direction-spin-commutator}
\end{aligned}
\end{equation}
This expression separates the spin-torque-like part from a purely
direction-space contribution. If the direction-space factors commute, or if
they are identical, the second term vanishes. In that case the commutator has
the spin structure used in the main text.

A particularly transparent case is obtained when both spin-dependent layers
enter with the same direction-space factor \(\bd\), with \(\bd^2=\one\). Then
\[
\Xmat_i=\gamma_i \bd\otimes(\vct m_i\cdot\bsig),
\]
and
\[
[\Xmat_2,\Xmat_1]
=
2\ii\gamma_1\gamma_2
\one_{\rm dir}\otimes
\left[
(\vct m_2\times\vct m_1)\cdot\bsig
\right].
\]
Thus the first commutator is compact in spin space and has the axial structure
associated with a field-like torque.

The next nested commutator gives
\[
[\Xmat_1,[\Xmat_2,\Xmat_1]]
=
-4\gamma_1^2\gamma_2
\bd\otimes
\left[
\vct m_1\times(\vct m_2\times\vct m_1)
\right]\cdot\bsig ,
\]
up to the sign convention used for the order of \(\Xmat_1\) and \(\Xmat_2\). This is
the vector structure associated with the damping-like component. Therefore the spin-only BCH expressions in the main text should be understood
as the common-direction-factor reduction of the full direction--spin
commutators.

For general direction-space factors \(\bd_i\), \eqref{eq:app-full-direction-spin-commutator}
shows that additional terms involving \([\bd_2,\bd_1]\) are generated. These terms
encode the coupling between spin structure and right/left-moving mode mixing.
We do not attempt a classification of all such terms.
	\bibliographystyle{apsrev4-1}
	\bibliography{bibliography}
	
\end{document}